\documentclass[lettersize,journal]{IEEEtran}
\usepackage{cite}
\usepackage{amsmath,amssymb,amsfonts}
\usepackage{algorithmic}
\usepackage{algorithm}
\usepackage{graphicx}
\usepackage{textcomp}
\usepackage{xcolor}
\usepackage{booktabs,hyperref}
\usepackage{multirow}
\usepackage{bm}
\usepackage{bbm}
\usepackage{subcaption}
\usepackage[dvipsnames]{xcolor}
\begin{document}

\title{\huge Efficient Quantum Algorithm for Phase Optimization of 1-Bit RIS-Assisted MIMO Communication System
}

\author{Soumyadip Paul and Neel Kanth Kundu
\thanks{
The work of Soumyadip Paul was supported in part by the IndiaAI Fellowship (App. ID.: PHD251015004118) under the IndiaAI Mission of the Ministry of Electronics and Information Technology (MeitY), Government of India. The work of Neel Kanth Kundu was supported in part by the INSPIRE Faculty Fellowship (Reg. No.: IFA22-ENG 344), ANRF Prime Minister Early Career Research Grant (ANRF/ECRG/2024/000324/ENS), and the New Faculty Seed Grant from the Indian Institute of Technology Delhi. This work was also supported by the University of Melbourne through the establishment of an IBM Quantum Network Hub.}
\thanks{
Soumyadip Paul is with the Center for Applied Research in Electronics (CARE), Indian Institute of Technology Delhi, New Delhi-110016, India (e-mail: crz258451@care.iitd.ac.in).


Neel Kanth Kundu is with CARE and the Bharti School of Telecommunication Technology and Management, Indian Institute of Technology Delhi, New Delhi-110016, India (e-mail: neelkanth@iitd.ac.in). He is also an honorary fellow at the Department of Electrical and Electronic Engineering, University of Melbourne, Melbourne, VIC-3010, Australia.}}

\maketitle
\begin{abstract}
We propose a Quantum Approximate Optimization Algorithm with a
deterministic linear ramp schedule (QAOA-LR) for phase optimization of 1-bit RIS-assisted 
MIMO communication system. Each RIS element is restricted to a binary
phase shift of $0$ or $\pi$, turning the passive beamforming design problem with $N$ elements into a
combinatorial optimization problem over $2^N$ configurations. Instead of running a
classical optimizer, QAOA-LR uses a fixed linear ramp to set the
variational parameters across $p$ layers and finds the best scale via a
simple one-dimensional grid search over a single parameter.
Monte Carlo simulations over Rayleigh-fading MIMO channels confirm that QAOA-LR closely tracks the optimum maximum-likelihood (ML) solution. Furthermore, the proposed algorithm reduces the time complexity in real hardware experiments on the IBM Quantum processor, confirming near ML capacity performance, with polynomial scaling of the quantum processing unit execution time as the number of RIS elements increases.
\end{abstract}


\section{Introduction}

Reconfigurable intelligent surfaces (RIS) have emerged as a transformative
paradigm for next-generation wireless communications, enabling software-controlled
electromagnetic beamforming at low hardware cost and power consumption\cite{8796365}.
By dynamically adjusting the phase and amplitude of incident signals, a passive RIS
can reshape the propagation environment without active radio-frequency chains,
making it a cost-effective complement to massive MIMO deployments. In practice,
hardware constraints often restrict each RIS element to a coarse 1-bit phase
resolution, $\phi_n \in \{0, \pi\}$, substantially reducing the circuit complexity
and control overhead\cite{11570206}. However, this binary quantization comes at an
algorithmic price: the resulting maximum likelihood (ML) phase optimization problem is combinatorial, with
$2^N$ candidate configurations for an $N$-element surface. This combinatorial
structure is in fact equivalent to the Max-Cut problem and is NP-hard in
general\cite{8811733}, ruling out exact exhaustive search beyond small $N$.
Classical approaches based on semidefinite programming relaxation (SDR) with
Gaussian randomization\cite{5447068,8811733} achieve near-optimal
performance in polynomial time and remain state-of-the-art for moderate $N$. However, the complexity of the SDR-based solutions with interior-point methods scales as $O(N^{3.5})$ per trial, which grows prohibitively
as the surface size scales to hundreds of elements envisioned in the 6G deployments \cite{5447068}. 


Recent advances in quantum computing have motivated the exploration of quantum-assisted techniques for addressing the combinatorial optimization challenges associated with RIS configuration. Quantum annealing (QA), implemented
on D-Wave hardware, maps the binary phase selection to a quadratic unconstrained
binary optimization (QUBO) problem and exploits quantum tunneling to escape local
minima\cite{11183656}. Prior work\cite{11183656} demonstrated
that QA with an augmented Lagrangian penalty scheme achieves superior capacity
performance over conventional heuristics for 1-bit RIS systems with index modulation, validating
the QUBO formulation on real annealing hardware. In the domain of gate-based
quantum computing, the Quantum Approximate Optimization
Algorithm (QAOA)\cite{farhi2014quantumapproximateoptimizationalgorithm} constructs
a parameterized quantum state through alternating cost and mixer unitaries that
has been studied for combinatorial graph problems such as
Max-Cut\cite{PhysRevA.97.022304,a12020034,PhysRevX.10.021067}. The work of
\cite{10838522} provides a broad survey of quantum algorithms for physical-layer
problems, while \cite{10602129,10829540} apply QAOA and related variational
methods for ML data detection in MIMO systems. More recently,
Colella \textit{et al.}\cite{11570206} applied QAOA to RIS beamforming
design and analyzed barren-plateau effects in the variational landscape. Furthermore,
a tutorial comparison of QA and QAOA for 1-bit RIS passive beamforming on
real quantum devices has been provided in\cite{11413851}, where
D-Wave annealing was found to deliver higher-quality solutions than low-depth QAOA under current hardware maturity. Collectively, these works confirm that quantum optimization is a compelling direction for RIS phase optimization, but they leave open critical scalability and practicality questions on gate-based hardware.


Despite these recent works, several important research gaps remain. First, standard QAOA relies on a classical outer-loop optimizer to tune its variational parameters, introducing significant parameter-optimization and repeated quantum-circuit evaluation overhead. Although linear-ramp parameterization has been investigated as an alternative to iterative QAOA parameter optimization\cite{l5r4-zcqv,MontanezBarrera2025LinearRampQAOA,Xiao2025QAOATruss,paul2026warmstartquantumapproximateoptimization}, its application to 1-bit RIS-assisted MIMO optimization has not been investigated yet. Second, existing QAOA-based treatments of RIS optimization are primarily limited to small MIMO configurations and simulation-based evaluations for small number of RIS elements, with results reported only up to $8\times8$ MIMO and relatively high circuit depths, while larger configurations such as $16\times16$ and $32\times32$ remain largely unexplored\cite{10747251}. Third, real-hardware validation and empirical evaluation of the computational overhead and execution-time behavior of QAOA-based RIS optimization remain limited, particularly for large-scale RIS configurations\cite{11570206,10747251}. This paper addresses these three gaps through the following contributions:
\begin{itemize}
\item We formulate the ML capacity problem for a 1-bit RIS-assisted MIMO system as an Ising Hamiltonian, incorporating pairwise coupling and single-qubit bias terms induced by direct-path cross-coupling, thereby enabling its implementation using QAOA.

\item We propose an efficient QAOA algorithm called QAOA-LR, which fixes parameters of the quantum circuit using a linear-ramp schedule and requires only a one-dimensional search over a single scale parameter. By eliminating iterative classical parameter optimization, the proposed approach reduces the associated optimization overhead.


\item We perform extensive Monte Carlo simulations with MIMO sizes upto $32 \times 32$ and up to $12$ RIS elements on classical simulators. Furthermore, we also validate the proposed QAOA-LR approach on IBM Quantum's \texttt{ibm\_marrakesh} processor for up to $100$ RIS elements and compare its runtime performance against classical baselines such as SDR and Randomized Restart Local Search (RRLS) methods.
\end{itemize}

The remainder of the paper is organized as follows. Section~\ref{sys_model} describes the RIS-assisted MIMO system model and the underlying ML phase-optimization problem. Section~\ref{sec:qaoa} formulates this problem as an Ising Hamiltonian and details the corresponding QAOA circuit construction. Section~\ref{sec:ramp} introduces the proposed QAOA-LR algorithm and its linear-ramp parameter schedule. Section~\ref{sec:complexity} analyzes the computational complexity of QAOA-LR relative to standard QAOA and classical baselines. Section~\ref{sec:sims} reports Monte Carlo simulation results together with real IBM Quantum hardware experiments, and Section~\ref{sec:conclusion} concludes the paper.

\section{System Model}\label{sys_model}

\begin{figure}[htp]
    \centering
\includegraphics[width=0.8\linewidth]{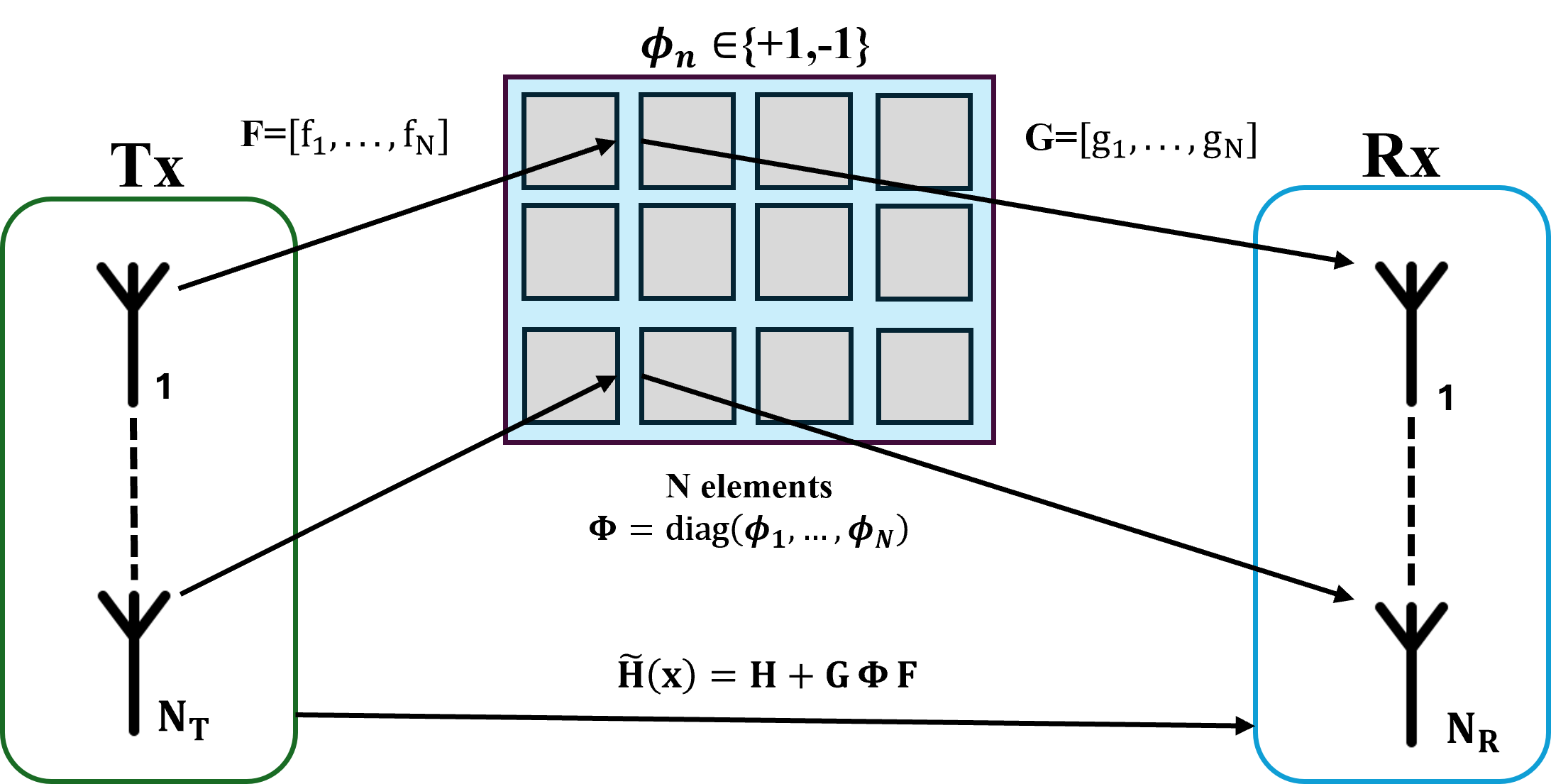}
    \caption{Schematic of the RIS-assisted MIMO communication system with $N_T$ transmit antennas, $N_R$ receive antennas,
    and a passive RIS with $N$ reflecting elements.
    }
    \label{fig:sysmodel}
\end{figure}

We consider a MIMO link assisted by a passive RIS with $N_T$ transmit antennas, $N_R$ receive antennas, and $N$ passive reflecting elements as shown in Fig.~ \ref{fig:sysmodel}. Each RIS element applies a 1-bit phase shift $\phi_n \in \{+1,-1\}$, collected into $\mathbf{x} = [\phi_1,\ldots,\phi_N]^T$ and $\boldsymbol{\Phi} = \mathrm{diag}(\mathbf{x})$ \cite{9110912}. The three wireless channels are denoted as (i) the direct channel between transmitter and receiver $\mathbf{H} \in \mathbb{C}^{N_R \times N_T}$, (ii) the RIS-to-receiver channel, $\mathbf{G} \in \mathbb{C}^{N_R \times N}$ and (iii) the transmitter-to-RIS channel, $\mathbf{F} \in \mathbb{C}^{N \times N_T}$ which are assumed to be i.i.d.\ $\mathcal{CN}(\mathbf{0},\mathbf{I})$ distributed under the Rayleigh fading model.
Neglecting higher-order reflections,
the received signal vector \(\mathbf{y} \in \mathbb{C}^{N_R \times 1}  \) is given by
\begin{equation}
    \mathbf{y} = \tilde{\mathbf{H}}(\mathbf{x})\mathbf{s} + \mathbf{z}
\end{equation}
where $\mathbf{s}$ denotes the transmitted signal with covariance $\mathbf{Q}$, $\mathrm{tr}(\mathbf{Q})\leq P_t$, with $P_t$ being the transmit power budget, $\mathbf{z}\sim\mathcal{CN}(\mathbf{0},\sigma^2\mathbf{I})$ denotes the additive Gaussian noise at the receiver, and $\tilde{\mathbf{H}}$ is
the effective MIMO channel matrix from the transmitter to the receiver  given by
\begin{equation}
    \tilde{\mathbf{H}}(\mathbf{x}) = \mathbf{H} + \mathbf{G}\boldsymbol{\Phi}\mathbf{F}, \quad
    ,
    \label{eq:heff}
\end{equation}

Under the assumption of perfect channel state information, the achievable capacity is given by\cite{9110912}
\begin{equation}
    C(\mathbf{x},\mathbf{Q})
    = \log_2\det\!\left(\mathbf{I}_{N_R} +
      \frac{1}{\sigma^2}\tilde{\mathbf{H}}(\mathbf{x})\mathbf{Q}\tilde{\mathbf{H}}(\mathbf{x})^H\right).
    \label{eq:cap}
\end{equation}
The optimal transmit covariance matrix is obtained via an SVD-based water-filling algorithm. However, at high SNR, an equal power allocation across all transmit antennas is near-optimal. Under this assumption, the covariance matrix simplifies to $\mathbf{Q} = \frac{P_t}{N_T}\mathbf{I}_{N_T}$, giving
\begin{equation}
    C(\mathbf{x})
    \approx \log_2\det\!\left(\mathbf{I}_{N_R} +
      \frac{P_t}{N_T \sigma^2}\tilde{\mathbf{H}}(\mathbf{x})\tilde{\mathbf{H}}(\mathbf{x})^H\right).
    \label{eq:capequal}
\end{equation}
Although this equal-power approximation is motivated by the high-SNR regime, our Monte Carlo results at $0$\,dB SNR confirm that maximizing $\|\tilde{\mathbf{H}}(\mathbf{x})\|_F^2$ closely tracks the true ML capacity even at moderate SNR, since the dominant source of capacity variation across phase configurations is the spread of singular values rather than the precise power allocation.

\section{QAOA Formulation for 1-Bit RIS Optimization} \label{sec:qaoa}

The ML phase optimization problem, maximizing the capacity, is given by
\begin{equation} \label{eq:ml}
    \mathbf{x}^{\mathrm{ML}} = \arg\max_{\mathbf{x}\in\{-1,+1\}^N} C^*(\mathbf{x}) \;.
\end{equation}
Note that the ML problem in \eqref{eq:ml} is NP-hard since we need to evaluate the objective for $2^N$ combinations of the binary phase shifts for $N$ RIS elements. To obtain a quantum-compatible objective, we exploit the monotone relationship between capacity and $\|\tilde{\mathbf{H}}(\mathbf{x})\|_F^2$. Expanding the Frobenius norm and dropping the constant term $\|\mathbf{H}\|_F^2$, the optimization reduces to maximizing
\begin{equation}
    f(\mathbf{x}) = \mathbf{x}^T \mathbf{R}\, \mathbf{x} + \mathbf{p}^T\mathbf{x},
    \label{eq:quad}
\end{equation}
where the coupling matrix $\mathbf{R} = \mathrm{Re}[\mathbf{F}^H\mathbf{G}^H\mathbf{G}\mathbf{F}] \in \mathbb{R}^{N\times N}$ captures pairwise RIS element interactions, and the bias vector $\mathbf{p} \in \mathbb{R}^N$ with $p_i = 2\,\mathrm{Re}[(\mathbf{H}^H\mathbf{G}\mathbf{F})_{ii}]$ encodes direct-path cross-coupling. When $\mathbf{H}$ is negligible, i.e, the line-of-sight link is heavily blocked, $\mathbf{p}=\mathbf{0}$ and the objective is purely quadratic.

Since QAOA solves a minimization problem, we negate\eqref{eq:quad} to obtain an equivalent problem given by
\begin{equation}
    \min_{\mathbf{x}\in\{-1,+1\}^N}\;
    C_{\mathrm{obj}}(\mathbf{x})
    = -\mathbf{x}^T\mathbf{R}\,\mathbf{x} - \mathbf{p}^T\mathbf{x}.
    \label{eq:cobj}
\end{equation}
Each spin $x_n \in \{-1,+1\}$ maps to a qubit via the Pauli-$Z$ eigenvalue correspondence $x_n \leftrightarrow \langle Z_n \rangle$, with $x_n{=}+1 \equiv |0\rangle_n$ and $x_n{=}-1 \equiv |1\rangle_n$. Classical products $x_i x_j$ become two-qubit operators $Z_i Z_j$. Substituting the spin-to-qubit map into\eqref{eq:cobj} and using $Z_n^2 = \mathbbm{1}$ to drop constant diagonal terms, yields the Ising Hamiltonian
\begin{equation}
    H_C
    = - 2\sum_{i<j} R_{ij}\, Z_i Z_j
      - \sum_{i=1}^N p_i\, Z_i,
    \label{eq:Hc_ising}
\end{equation}
with pairwise couplings $J_{ij} = -2R_{ij}$ and local fields $h_i = -p_i$. Its ground state solves\eqref{eq:quad}.

\begin{figure}[htp]
    \centering
    \includegraphics[width=1\linewidth]{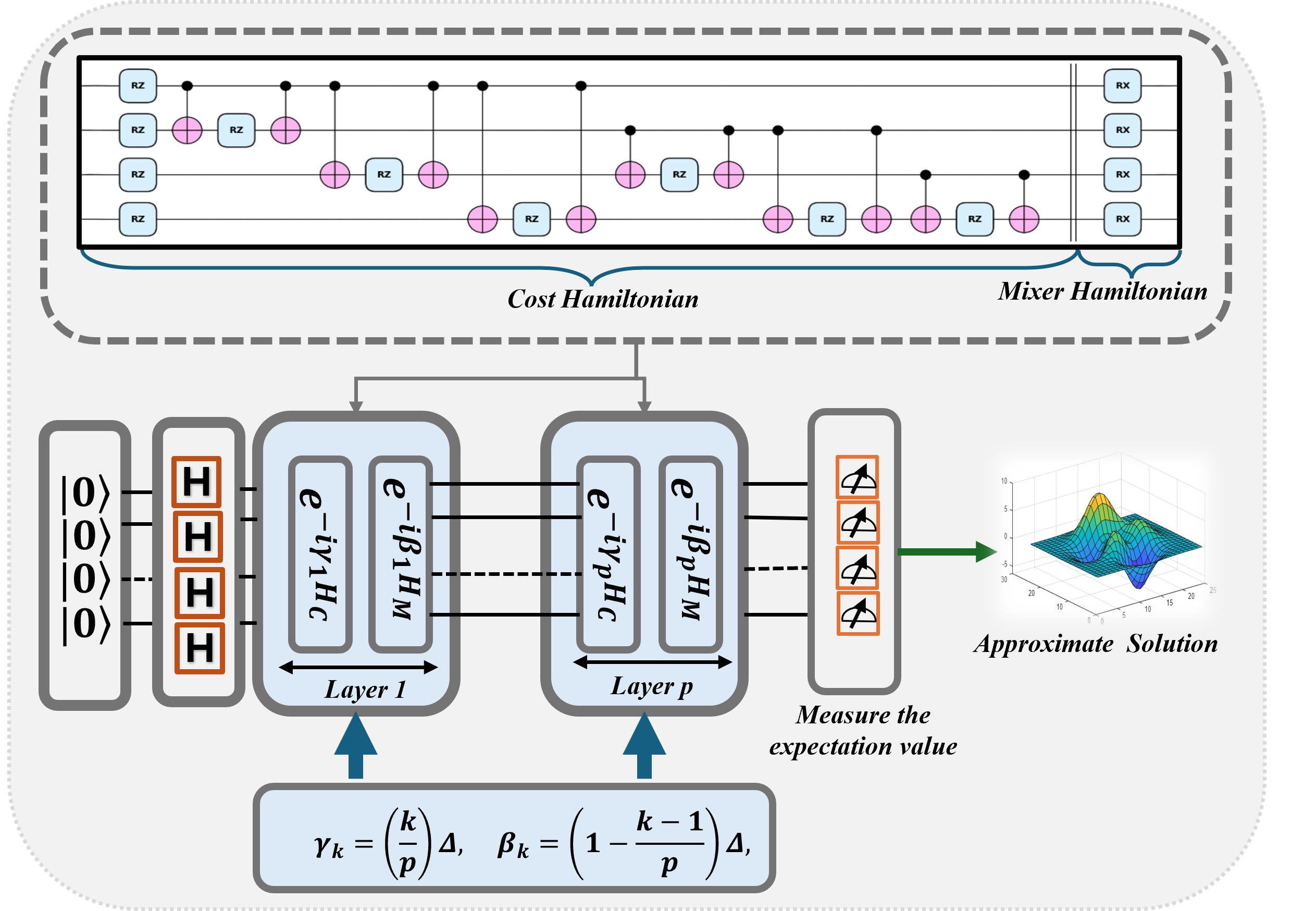}
    \caption{QAOA-LR circuit for 1-bit RIS phase optimization.}
    \label{fig:circuit}
\end{figure}

\begin{figure}
    \centering
    \includegraphics[width=0.75\linewidth]{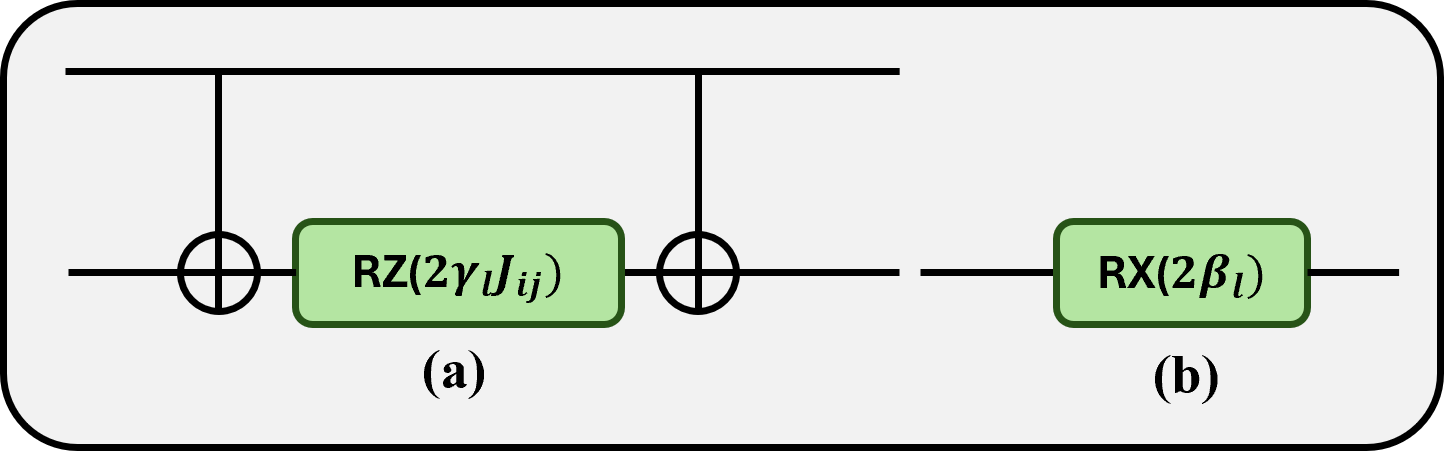}
    \caption{Quantum gates. (a) Decomposition of the $Z_iZ_j$ coupling term into a $\mathrm{CNOT}$--$\mathrm{RZ}(2\gamma_\ell J_{ij})$--$\mathrm{CNOT}$ sequence. (b) Single-qubit $\mathrm{RX}(2\beta_\ell)$ rotation used in the mixer unitary.}
    \label{fig:gates}
\end{figure}

An equivalent QUBO form can be obtained by substituting $x_n = 1-2b_n$, $b_n\in\{0,1\}$, which maps each spin variable onto a binary decision variable. This is the standard reduction used to express Ising-model ground-state problems as QUBO instances for solvers that operate natively on binary $\{0,1\}$ variables rather than $\{-1,+1\}$ spins. Let $\mathbf{W}$ denote the QUBO matrix given by
\begin{equation}
    \mathbf{W}
    = -4\mathbf{R}
      + 4\,\mathrm{diag}(\mathbf{R}\,\mathbf{1}_N)
      - 2\,\mathrm{diag}(\mathbf{p}),
    \label{eq:Wmat}
\end{equation}
where $\mathbf{1}_N \in \mathbb{R}^N$ 
denotes the all-ones column vector and $\mathrm{diag}(\mathbf{R}\mathbf{1}_N)$ is the $N\times N$ diagonal matrix whose $i$-th diagonal entry equals $\sum_{j=1}^N R_{ij}$. Using \eqref{eq:Wmat}, the cost Hamiltonian can be expressed as
\begin{equation}
    H_C
    = \sum_{i} \frac{W_{ii}}{4}\,Z_i
      + \sum_{i<j} \frac{W_{ij}}{4}\,Z_i Z_j.
    \label{eq:Hc_qubo}
\end{equation}
The transverse-field mixer $H_M = \sum_{n=1}^{N} X_n$ generates quantum tunneling between bitstring configurations. Starting from the initial quantum state $|\psi_{\mathrm{init}}\rangle = |+\rangle^{\otimes N}$, a depth-$p$ circuit prepares
\begin{equation}
    |\psi(\boldsymbol{\gamma},\boldsymbol{\beta})\rangle
    = \prod_{\ell=1}^{p}
      e^{-i\beta_\ell H_M}
      e^{-i\gamma_\ell H_C}
      |\psi_{\mathrm{init}}\rangle,
    \label{eq:qaoa_state}
\end{equation}
with the parameters $\boldsymbol{\gamma} = [\gamma_1,\gamma_2,\ldots,\gamma_p], \boldsymbol{\beta} = [\beta_1,\beta_2,\ldots,\beta_p]$ fixed by the linear ramp method described in Section \ref{sec:ramp}. The resulting layered structure of the QAOA-LR circuit, which alternates the cost unitary $e^{-i\gamma_\ell H_C}$ and the mixer unitary $e^{-i\beta_\ell H_M}$ over $p$ repetitions starting from $|\psi_{\mathrm{init}}\rangle$, is depicted in Fig.~\ref{fig:circuit}. The cost unitary $e^{-i\gamma_\ell H_C}$ uses single-qubit $\mathrm{RZ}(2\gamma_\ell h_i)$ gates for the local-field terms and $\mathrm{CNOT}$--$\mathrm{RZ}(2\gamma_\ell J_{ij})$--$\mathrm{CNOT}$ sequences for the $Z_iZ_j$ coupling terms. The mixer unitary $e^{-i\beta_\ell H_M}$ decomposes into single-qubit rotations $\mathrm{RX}(2\beta_\ell)$ applied independently to every qubit $n \in \{1,\ldots,N\}$, i.e., $e^{-i\beta_\ell H_M} = \bigotimes_{n=1}^{N} \mathrm{RX}_n(2\beta_\ell)$ as shown in Fig.\ref{fig:gates}. The solution is decoded as
\begin{equation}
    x_n^{\mathrm{QAOA}}
    = \mathrm{sign}\!\left(\langle Z_n\rangle\right),
    \label{eq:decode}
\end{equation}
with ties rounded to $+1$.

\section{QAOA with Linear Ramp Schedule Algorithm} 
\label{sec:ramp}


Rather than running a classical optimizer over the non-convex QAOA energy landscape, we use a linear ramp schedule inspired by adiabatic quantum computing. For layer $k \in \{1, \ldots, p\}$, the parameters are set as
\begin{equation}
    \gamma_k = \frac{k}{p}\,\Delta,
    \qquad
    \beta_k  = \left(1 - \frac{k-1}{p}\right)\Delta,
    \label{eq:ramp}
\end{equation}
where $\Delta > 0$ controls the overall parameter scale\cite{paul2026warmstartquantumapproximateoptimization}. Early layers favor exploration (large $\beta_k$, small $\gamma_k$), while later layers drive toward low-cost states. We select $\Delta$ by evaluating the circuit over a discrete grid and picking the value minimizing the expected cost $\mathcal{E}(\boldsymbol{\gamma},\boldsymbol{\beta}) = \langle\psi|H_C|\psi\rangle$, making QAOA-LR fully independent of a classical variational optimizer and well suited to noisy hardware. The complete procedure is outlined in Algorithm\ref{alg:qaoa_lr}.

\begin{algorithm}[htp]
\caption{QAOA with Linear Ramp Schedule (QAOA-LR)}
\label{alg:qaoa_lr}
\begin{algorithmic}[1]
\REQUIRE Channel matrices $\mathbf{F}$, $\mathbf{G}$, $\mathbf{H}$; QAOA depth $p$; grid of scale parameters $\Delta_{\mathrm{grid}}$
\ENSURE  Optimum binary phase vector $\hat{\mathbf{x}}$ and capacity $\hat{C}$
\STATE Compute $\mathbf{R} = \mathrm{Re}[\mathbf{F}^H\mathbf{G}^H\mathbf{G}\mathbf{F}]$ and $p_i = 2\mathrm{Re}[(\mathbf{H}^H\mathbf{G}\mathbf{F})_{ii}]$
\STATE Formulate QUBO: $\mathbf{W} = -4\mathbf{R} + 4\mathrm{diag}(\mathbf{R}\mathbf{1}_N) - 2\mathrm{diag}(\mathbf{p})$\STATE Initialize $\mathcal{E}^* \leftarrow \infty$, $\hat{\Delta} \leftarrow 0$
\FOR{$\Delta \in \Delta_{\mathrm{grid}}$}
    \FOR{$k = 1$ \TO $p$}
        \STATE $\gamma_k \leftarrow \frac{k}{p}\Delta$, \quad $\beta_k \leftarrow \left(1 - \frac{k-1}{p}\right)\Delta$
    \ENDFOR
    \STATE Execute QAOA circuit; measure $\mathcal{E}(\Delta) = \langle\psi| H_C |\psi\rangle$
    \IF{$\mathcal{E}(\Delta) < \mathcal{E}^*$}
        \STATE $\mathcal{E}^* \leftarrow \mathcal{E}(\Delta)$, $\hat{\Delta} \leftarrow \Delta$; save $\langle Z_n \rangle$
    \ENDIF
\ENDFOR
\STATE $x_n^{\mathrm{QAOA}} \leftarrow \mathrm{sign}(\langle Z_n \rangle)$ for all $n$
\STATE $\hat{C} \leftarrow \log_2\det\left(\mathbf{I} + \frac{P_t}{N_0 N_T}\mathbf{H}_{\mathrm{eff}}(\hat{\mathbf{x}})\mathbf{H}_{\mathrm{eff}}(\hat{\mathbf{x}})^H\right)$
\RETURN $\hat{\mathbf{x}}$, $\hat{C}$
\end{algorithmic}
\end{algorithm}

\begin{table*}[t]
\centering
\caption{Mean capacity (bits/s/Hz) for ML and QAOA-LR across MIMO configurations.}
\label{tab:capacity}
\begin{tabular}{|c|cc|cc|cc|cc|cc|}
\toprule
\bottomrule
&  \multicolumn{2}{c|}{$2{\times}2$} & \multicolumn{2}{c|}{$4{\times}4$} & \multicolumn{2}{c|}{$8{\times}8$} & \multicolumn{2}{c|}{$16{\times}16$} & \multicolumn{2}{c}{$32{\times}32$} \\
$N$  & ML & QAOA-LR & ML & QAOA-LR & ML & QAOA-LR & ML & QAOA-LR & ML & QAOA-LR \\
\bottomrule
\midrule
1  & 2.879 & 2.879 & 4.997 & 4.997 & 8.874 & 8.874 & 16.452 & 16.452 & 30.676 & 30.676 \\
2   & 3.975 & 3.975 & 6.425 & 6.425 & 10.670 & 10.670 & 19.381 & 19.381 & 34.224 & 34.224 \\
4   & 5.431 & 5.431 & 8.736 & 8.730 & 14.669 & 14.654 & 24.878 & 24.878 & 41.816 & 41.816 \\
6   & 6.401 & 6.378 & 10.248 & 10.182 & 17.538 & 17.536 & 29.266 & 29.266 & 48.557 & 48.557 \\
8  & 8.113 & 8.059 & 12.455 & 12.399 & 20.141 & 20.110 & 34.018 & 34.018 & 55.713 & 55.713 \\
9   & 8.276 & 8.160 & 13.235 & 13.203 & 21.714 & 21.680 & 36.739 & 36.684 & 58.638 & 58.638 \\
10  & 8.882 & 8.646 & 13.815 & 13.701 & 22.939 & 22.894 & 37.594 & 37.594 & 61.379 & 61.378 \\
12  & 9.547 & 9.295 & 14.973 & 14.776 & 24.932 & 24.880 & 42.355 & 42.355 & 68.028 & 68.028 \\
\midrule
\bottomrule
\end{tabular}
\end{table*}
\begin{figure}[!hpb]
    \centering
    \begin{subfigure}[t]{0.8\linewidth}
        \centering
    \includegraphics[width=\linewidth]{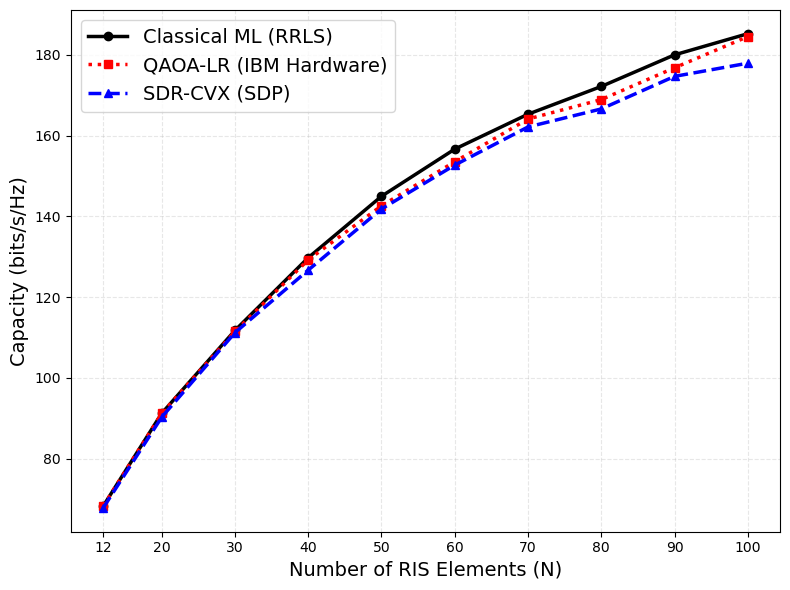}
    \caption{Capacity vs Elements\ $N$}
    \label{fig:hw_capacity}
    \end{subfigure}
    \hfill
    \begin{subfigure}[t]{0.8\linewidth}
        \centering
        \includegraphics[width=\linewidth]{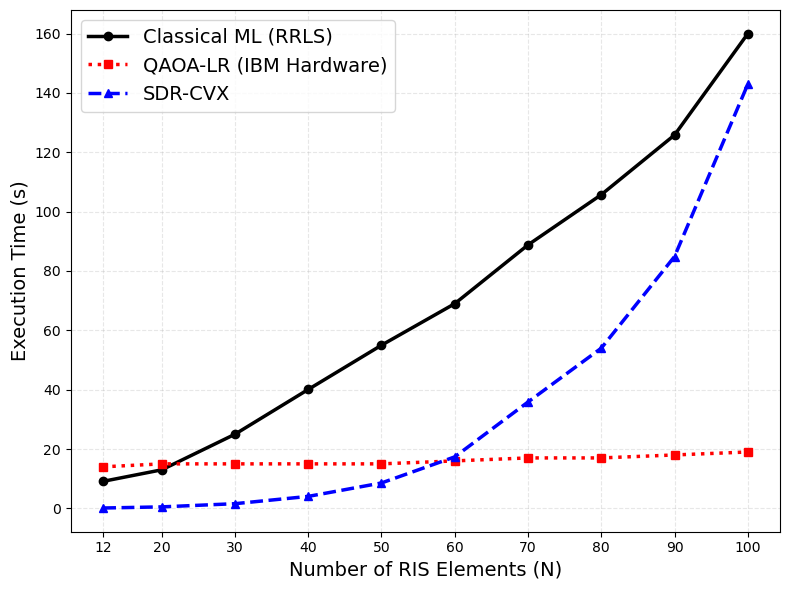}
    \caption{Execution time vs.\ $N$.}
    \label{fig:hw_runtime}
    \end{subfigure}
    \begin{subfigure}[t]{0.8\linewidth}
        \centering
        \includegraphics[width=\linewidth]{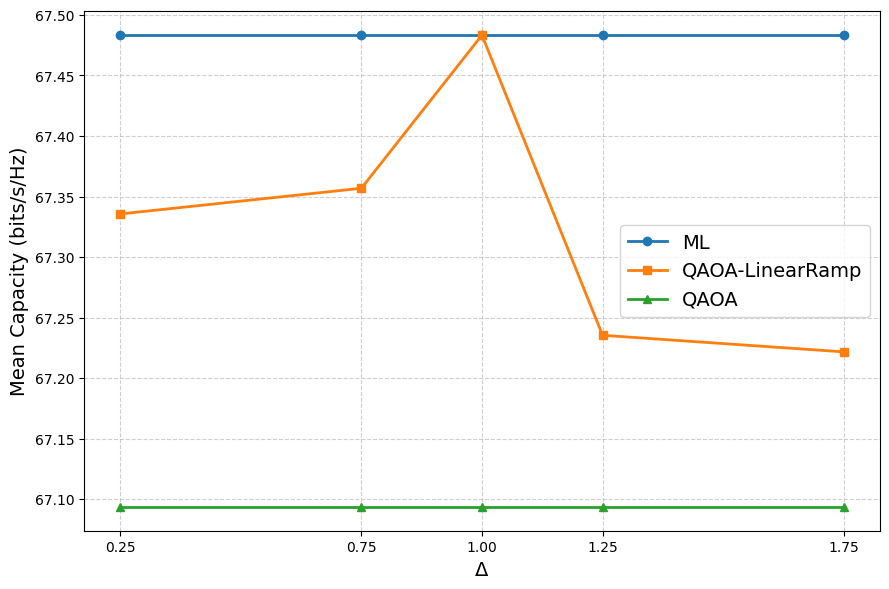}
    \caption{Mean capacity versus the linear-ramp scale parameter $\Delta$ for an $N=12$.}
    \label{fig:del}
    \end{subfigure}
    \caption{Real-hardware evaluation on IBM Quantum's \texttt{ibm\_marrakesh} processor for a $32{\times}32$ MIMO-RIS system with $N \in \{12,\ldots,100\}$: (a) capacity, (b) execution time, and (c) sensitivity to the linear-ramp scale parameter $\Delta$.}
\end{figure}

\begin{table}[t]
\centering
\caption{Complexity comparison.}
\label{tab:complexity}
\begin{tabular}{|c|c|c|c|}
\toprule
 \textbf{ML} &  \textbf{SDR} & \textbf{QAOA-LR} & \textbf{QAOA } \\
\midrule
 $O(2^N)$ & $O(N^{3.5})$ & $O(MpN)$ & $O(dpN)$ \\
\bottomrule
\bottomrule
\end{tabular}
\end{table}

\section{Complexity Analysis} \label{sec:complexity}
The QAOA-LR circuit requires $O(pN^2)$ gate operations per grid point for a dense Ising Hamiltonian. Unlike conventional QAOA, which iteratively optimizes a $2p$-dimensional parameter vector through repeated quantum-circuit evaluations, QAOA-LR parameterizes all angles using a single scale parameter $\Delta$ and performs an $M$-point grid search. Hence, its overall gate complexity is $O(MpN^2)$. Exhaustive ML search requires $O(2^N)$ evaluations, whereas the SDR approach has polynomial complexity $O(N^{3.5})$\cite{5447068}. Under the adopted hardware execution-time model, QAOA-LR requires $O(MpN)$\cite{Xiao2025QAOATruss,crooks2018performancequantumapproximateoptimization}, while conventional QAOA incurs $O(dpN)$ complexity for $d$ ($d>M$) classical optimization iterations. Thus, QAOA-LR replaces the iterative $2p$-dimensional parameter optimization of conventional QAOA with a one-dimensional search over $\Delta$, thereby reducing the classical parameter-optimization overhead. The time-complexity comparison is summarized in Table ~\ref{tab:complexity}.

\section{Simulation Results} \label{sec:sims}


We evaluate QAOA-LR with $N \in \{2,4,6,8,9,10,12\}$ RIS elements and
six MIMO configurations ($N_T{=}N_R \in \{2,4,8,16,32\}$). All channel matrices are drawn i.i.d.\ from $\mathcal{CN}(\mathbf{0},\mathbf{I})$  with transmit SNR $P_t/\sigma^2 = 1$ (0\,dB), and capacity is evaluated via \eqref{eq:capequal} over 5000 Monte Carlo trials. The QAOA circuit uses depth $p=8$ and grid of linearly spaced $\Delta$ values in $[0.25,1.75]$. The mean capacity achieved by ML and QAOA-LR across the five MIMO configurations and for RIS element counts $N$ from 2 to 12 is summarized in Table \ref{tab:capacity}. It can be observed that for $N \leq 2$, the capacity performance of QAOA-LR matches exactly with that of the ML method across all MIMO configurations.
As $N$ increases to 12, a slight gap appears for smaller antenna arrays, but the gap shrinks
consistently as antenna size grows. For $32{\times}32$ MIMO, the performance of QAOA-LR closely follows the
ML at every tested $N$, confirming the deterministic
linear ramp as an effective and robust heuristic for the QAOA-based RIS phase optimization framework.

Next, we ran experiments on IBM Quantum's \texttt{ibm\_marrakesh} processor with
1000 shots per circuit, fixing a $32{\times}32$ MIMO configuration and varying $N$ from 12 to 100. Owing to limited QPU access time and queue resources, we did not repeat the $\Delta$ grid search on hardware; instead, each trial submits a single circuit-execution job at a fixed $\Delta{=}1.0$ selected from the classical-simulation results, with circuit depth $p{=}6$, followed by a final decoding job. Since exhaustive ML search becomes intractable for $N \geq 20$, we use Randomized Restart Local Search (RRLS) \cite{doi:10.1137/15M1047775} as the classical baseline in place of true ML for the capacity comparison, together with an SDR baseline and the proposed QAOA-LR. 

The mean capacity attained by RRLS, SDR, and QAOA-LR as a function of the number of RIS elements $N$ is compared in Fig.\ref{fig:hw_capacity}. At $N{=}12$, the real quantum processor achieves $68.2366$\,bits/s/Hz, matching the ideal simulation, and QAOA-LR continues to deliver high capacity up to $N{=}100$, confirming its scalability on NISQ devices. The corresponding execution-time behavior of RRLS, SDR-CVX\cite{9198125}, and QAOA-LR as $N$ grows is compared in the runtime plot shown in Fig.\ref{fig:hw_runtime}. The QPU firmware time for QAOA-LR scales linearly with $N$, while the RRLS and SDR-CVX runtimes both grow superlinearly, highlighting the favorable hardware execution-time scaling of the proposed approach. Next, we examine the sensitivity of QAOA-LR to its single scale parameter. The mean capacity versus the linear-ramp scale parameter $\Delta$ for $N{=}12$ in a $32\times32$ MIMO configuration, swept over the grid of linearly spaced values and compared against Adam-optimized QAOA, is shown in Fig.~\ref{fig:del}. QAOA-LR peaks at $\Delta{=}1.0$ and outperforms the Adam baseline across all $\Delta$, confirming that a one-dimensional grid search suffices without iterative classical optimization.

\section{Conclusion} \label{sec:conclusion}
We have proposed an efficient quantum optimization algorithm called QAOA-LR for the phase design of 1-bit RIS-assisted MIMO communication systems. The phase optimization problem is mapped to an Ising Hamiltonian, and the iterative classical parameter optimization of conventional QAOA is replaced by a one-dimensional search over the linear-ramp scale parameter. Monte Carlo simulations for MIMO configurations from $2{\times}2$ to $32{\times}32$ demonstrate near-optimal capacity performance. Furthermore, real-hardware experiments on IBM Quantum's \texttt{ibm\_marrakesh} processor with up to $100$ RIS elements validate the practical feasibility and favorable execution-time behavior of the proposed approach. Future work will extend the proposed framework to multi-bit RIS phase optimization.

\bibliographystyle{IEEEtran}
\bibliography{references}

\end{document}